\documentstyle[12pt]{article}
\def\a{\alpha}\def\b{\beta}\def\g{\gamma}\def\d{\delta}
\def\E{\varepsilon}
\def\l{\lambda}\def\s{\sigma}
\def\om{\omega}\def\G{\Gamma}
\def\semiprod{\subset\!\!\!\!\!\!\times}
\newcommand{\p}[1]{(\ref{#1})}

\begin{document}
\renewcommand{\thefootnote}{\fnsymbol{footnote}}
\thispagestyle{empty}
\def\theequation{\thesection.\arabic{equation}}

\begin{flushright}
{\bf FTUV/98-97, IFIC/98-99, \\ 
TUW-98-25 ,\\
hep-th/9812074} \\
 December 9, 1998
\end{flushright}

\vskip0.5cm

\begin{center}
{\Large \bf New Superparticle Models \\ 
Outside the HLS Supersymmetry Scheme}\footnote{
To be published in the Proceedings of 12-th Max Born Symposium: 
Theoretical Physics - Fin de Siecle, held 24-27.09.1998 in Wroclaw 
(Poland), ed. by A. Borowiec, B. Jancewicz, W. Karwowski and 
W. Cegla,  Springer Verlag. }

\bigskip

{\large \bf Igor Bandos$^{1)}$\footnote{ Lise Meitner Fellow.
On leave of absence from  Institute for Theoretical Physics,
NSC Kharkov Institute of Physics and Technology,
310108 Kharkov, Ukraine}
 and Jerzy Lukierski}$^{2)}$\footnote{
On leave of absence from 
Institute for Theoretical Physics, 
University of Wroclaw, 
50-204 Wroclaw, Poland.  
Supported in part by {\bf KBN} grant
{\bf 2P03B13012}
}
 \\
$^{1)}$ {\it Institute for Theoretical Physics\\
Technical University of Vienna\\
Wiedner Haupstrasse 8-10\\
A-1040 Wien, Austria}\\
{\bf e-mail: bandos@tph32.tuwien.ac.at}
 \\
$^{2)}${\it 
Departamento Di Fisica Teorica \\
University of Valencia \\ 
46300 Burjasot (Valencia), Spain,} \\
{\bf e-mail: lukier@condor1.ific.uv.es}
\date{}
\end{center}

\vskip0.3cm


\begin{abstract}
{\small 
We consider the superparticle models invariant under the supersymmetries 
with tensorial central charges, which were not included in $D=4$ 
Haag-Lopuszanski-Sohnius (HLS) supersymmetry scheme. 

We present  firstly 
a  generalization of $D=4$ 
Ferber-Shirafuji (FS) model with fundamental
bosonic spinors and tensorial central charge coordinates.
The model contains four fermionic
coordinates and possesses three $\kappa$-symmetries thus providing
the BPS configuration preserving $3/4$ of the target space supersymmetries.
We show that the
physical degrees of freedom ($8$ real bosonic and $1$ real
Grassmann variable) of our
model can be described by $OSp(8|1)$ supertwistor.
Then 
we propose  a higher dimensional generalization of our model
with one real fundamental bosonic spinor.
$D=10$  model describes massless superparticle with
composite tensorial central charges and in $D=11$ we obtain 0-superbrane
model with  nonvanishing mass which is generated dynamically.
The introduction of $D=11$ Lorentz harmonics provides the possibility 
to construct 
massless $D=11$ superparticle model which can be formulated in a way 
preserving $1/2$, $17/32$, $18/32$, $\ldots$, $31/32$ supersymmetries. 
In a special case  we obtain 
the twistor-like formulation of the usual massless
$D=11$ superparticle proposed recently by Bergshoeff and Townsend. 
}

\end{abstract}

\setcounter{page}0
\renewcommand{\thefootnote}{\arabic{footnote}}
\setcounter{footnote}{0}
\newpage

\section{Introduction}
\setcounter{equation}{0}

It is our great pleasure to contribute this article to the volume
dedicated to Professor Jan Lopuszanski on his 75-th birthday. He is one of
the founders of algebraic background for present supersymmetric theories.
In seventies, when in 1975 he published fundamental paper with Haag and
Sohnius (see \cite{HLS}) it was however assumed that the relativistic 
superalgebra should contain in its bosonic sector a direct summ of
 space-time symmetry generators ( Poincar\'{e}, de-Sitter, conformal) and
internal symmetry generators, i.e. the space-time bosonic generators and 
internal bosonic generators should commute. As a consequence the internal
Abelian generators, called also central charges, had to be scalar.
Recently however this conclusion has been relaxed, and in present
algebraic framework of SUSY appear generalized central charges - tensorial
\cite{HP}--\cite{Hewson} or even spinorial \cite{AF,S} ones. 
The best example can be provided by
D=11 supersymmetry algebra, containing topological contributions from
M2 and M5 superbranes:
\begin{equation}\label{QQZ}
\{ Q_{\alpha}, Q_{\beta} \} = 
 P_m \G^m_{\a\b}+
Z_{{m}_1{m}_2} \G^{{m}_1{m}_2}
_{\b\a}+ Z_{m_1...m_5}
\G^{m_1...m_5}_{\b\a} . 
\end{equation}

 In this lecture we shall consider the new superparticle models,
invariant under SUSY with tensor charge generators. We shall formulate
such a model following the ideas of supertwistor formulation by Ferber
and Shirafuji 
\cite{F78,S83}. In Sect 2 we shall consider the D=4 model which is
invariant under the following D=4 SUSY algebra
\begin{equation}\label{QQZ4}
\{ Q_A, Q_B \} = Z_{AB} , \qquad
\{ \bar{Q}_{\dot{A}}, \bar{Q}_{\dot{B}} \} = \bar{Z}_{\dot{A}\dot{B}} , \qquad
\end{equation}
$$
\{ Q_A, \bar{Q}_{\dot{B}} \} = P_{A\dot{B}} , \qquad
$$
where $ (Q_A)^* = \bar{Q}_{\dot{A}}$,
$(P_{A\dot{B}})^* = P_{B\dot{A}}$,
$(Z_{{A}{B}})^* =
\bar{Z}_{\dot{A}\dot{B}}$ and six real
commuting central charges $Z_{\mu\nu}= - Z_{\nu\mu}$ are related to the
symmetric complex
spin-tensor $Z_{AB}$ by
\footnote{For two-component $D=4$ Weyl spinor formalism see e.g. \cite{OM75}.
We have \\ $(\s_{mn})_A^{~B}= {1\over 2i} \big(
(\s_{\mu} )_{A\dot{B}}
\tilde{\s}_{\nu}^{\dot{B}B}-
(\s_{\nu} )_{A\dot{B}}
\tilde{\s}_{\mu}^{\dot{B}B}\big)=
 - {i \over 2} \E_{\mu\nu\rho\l}(\s^{\rho\l})_A^{~B} =
[(\tilde{\s}_{\mu\nu})^{\dot{B}}_{~\dot{A}}]^* $. }
\begin{equation}\label{ZZZ}
Z_{\mu\nu} = {i \over 2}\left( \bar{Z}_{\dot{A}\dot{B}}
\tilde{\s}_{\mu\nu}^{\dot{A}\dot{B}}
- Z_{AB} \s_{\mu\nu}^{AB}\right).
\end{equation}
Thus the spin-tensors $Z_{AB}$ and $Z_{\dot{A}\dot{B}}$
$$
Z_{AB} ={i \over 4} Z_{\mu\nu} \s^{\mu\nu}_{AB}, \qquad
\bar{Z}_{\dot{A}\dot{B}} =
 - {i \over 4 }{Z}_{\mu\nu} \tilde{\s}^{\mu\nu}_{\dot{A}\dot{B}}
$$
represent the self-dual and anti-self-dual parts of the central charge
matrices.
It should be stressed that the superalgebra (1.2-3) goes outside of the 
HLS scheme.

The $D=4$ model considered in Section 2 can be reformulated in terms of two Weyl spinors
$\l_A, \mu_A$ and one real Grassmann variable $\zeta$
expressed by the
generalization
of supersymmetric  Penrose--Ferber
relations \cite{F78,S83,BBCL} between supertwistor and superspace coordinates.
  Such reformulation is described by $OSp(8|1)$ invariant free supertwistor
  model with the action
\begin{equation}
 \label{act0}
S = - {1 \over 2} \int d \tau {Y}_{{\cal A}} G^{{\cal A} {\cal B}}
\dot{Y}_{{\cal B}}
\end{equation}
where ${Y}_{{\cal A}} = (y_1, \ldots , y_8; \zeta)
\equiv
(\l_\a , \mu^\a, \zeta )$ is the real $SO(8|1)$ supertwistor
  (see e.g. \cite{Lstw}) and
\begin{equation}
 \label{OSpm}
  G^{{\cal A}{\cal B}} = \left( \matrix{
   \om^{(8)} & 0 \cr
   0 & 2i \cr} \right) =
   \left(
  \matrix{
  \left(\underline{\matrix{
   0_2 & I_2 & 0_2 & 0_2  \cr
  -I_2 & 0_2 & 0_2 & 0_2   \cr
   0_2 & 0_2 & 0_2 & I_2   \cr
  0_2 & 0_2 & {-I}_2 & 0_2 \cr  }}\right)
    & |& 0   \cr
   0 & |& i \cr}           \right)
\end{equation}
 is the $OSp(8|1)$ supersymplectic structure with bosonic $Sp(8)$ symplectic
 metric \\ $\om^{(8)}=-(\om^{(8)})^T$.
It should be mentioned therefore that due to the presence of
 tensorial central charges the standard $SU(2,2|1)$ supertwistor description
 \cite{F78,S83,BBCL,stv,SorG,Town91}
 of the Brink--Schwarz (BS) massless superparticle \cite{BS}
 with one complex Grassmann coordinate
 is replaced by a model with $OSp(8|1)$ invariance and one real Grassmann degree of
 freedom.

It should be stressed that by the use of spinor coordinates in the presence of tensorial central charges

\begin{itemize}
\item
we do not increase the initial number of spinor degrees of freedom
(four complex or eight real components)
in comparison with the model without
tensorial central charges;
\item we keep the manifest Lorentz invariance despite the presence of
tensorial central charges.
\end{itemize}

In fact, when we use our formulae (see Section {\bf 3})
 \begin{equation}\label{CPrep}
P_{A\dot{B}} = \lambda_{A}\bar{\lambda}_{\dot{B}},
\qquad
Z_{AB} = \lambda_{A} \lambda_{B},
\qquad
\bar{Z}_{\dot{A}\dot{B}} = \bar{\lambda}_{\dot{A}}
\bar{\lambda}_{\dot{B}}
\end{equation}
we find
 that, in comparison with standard FS model
 ($P_{A\dot{B}} = \l_A\bar{\l}_{\dot{B}}$, 
$Z_{AB}=\bar{Z}_{\dot{A}\dot{B}}=0$),
 only the phase of
 spinor $\l_A$ becomes an additional physical bosonic degree of freedom.

\bigskip 

In Section 3 we shall consider the D=10 and D=11 models described by 
multidimensional extensions of FS model with one fundamental spinor 
coordinates. The  D=11 model is invariant under
the superalgebra \p{QQZ}.
It appears that D=10 model is massless  (due to the famous Fierz 
identities for D=10 gamma matrices) 
and D=11 is generally a massive one with a mass generated dynamically. 
In Section 4 we shall consider the large family of D=11 massless models 
with particular fundamental spinor coordinates described by Lorentz 
harmonics. 

We would like to add that the results presented in Sections 2 and 3 can 
also be found in our recent article \cite{BL}, but all the results from 
Section 4 are new.

\bigskip

\section{ Generalization of Ferber--Shirafuji superparticle model:
spinor fundamental variables and central charges}
\setcounter{equation}{0}

 We generalize the model presented in \cite{S83} as follows
\begin{equation}\label{action}
S = \int d \tau
\left( \lambda_{A}\bar{\lambda}_{\dot{B}} \Pi_\tau^{A\dot{B}} +
\lambda_{A}\lambda_{B}  \Pi_\tau^{AB}
+ \bar{\lambda}_{\dot{A}} \bar{\lambda}_{\dot{B}}
\, \Pi_\tau^{\dot{A}\dot{B}} \right) \qquad ,
\end{equation}
where
\begin{equation}\label{vielbeine}
\begin{array}{l}
\displaystyle
\Pi^{A\dot{B}} \equiv  d\tau \Pi_\tau^{A\dot{B}}
= d{X}^{A\dot{B}}
+ i \left( d\Theta^{A} \bar{\Theta}^{\dot{B}}
- \Theta^{A} d\bar{\Theta}^{\dot{B}}\right) \, ,
\\ \nonumber
\displaystyle
\Pi^{AB} \equiv d\tau \Pi_\tau^{AB} =
d{z}^{AB} - ~i~ \Theta^{(A}~d{\Theta}^{B)}\, ,
\\  
\displaystyle
\bar{\Pi}^{\dot{A}\dot{B}}
\equiv d\tau \bar{\Pi}_\tau^{\dot{A}\dot{B}}
= d{\bar{z}}^{\dot{A}\dot{B}}
- ~i ~\bar{\Theta}^{(\dot{A}} ~d{\bar{\Theta}}\,{}^{\dot{B})}\, ,
\end{array}
\end{equation}
are the supercovariant one--forms in $D=4$, $N=1$ generalized flat superspace
\begin{equation}\label{superspace}
M^{(4+6|4)} = \{ Y^M \} \equiv \{ (X^{A\dot{A}}, z^{AB},
\bar{z}^{\dot{A}\dot{B}}; \Theta^{A}, \bar{\Theta}^{\dot{A}})\},
\end{equation}
{\sl with tensorial central charge coordinates}
 $z^{mn} = (z^{AB}, \bar{z}^{\dot{A}\dot{B}} )$ (see \p{ZZZ}).
The complete configuration space of the model \p{action}  contains
 additionally the complex-conjugate pair $(\l_A, \bar{\l}_{\dot{A}})$
 of Weyl spinors                      
\begin{equation}\label{confsuperspace}
{\cal M}^{(4+6+4|4)} = \{ q^{{\cal M}} \} \equiv
\{  (Y^M; \l^A, \bar{\l}^{\dot{A}}) \}
=\{ (X^{A\dot{A}}, z^{AB},
\bar{z}^{\dot{A}\dot{B}};  \l^A, \bar{\l}^{\dot{A}};
\Theta^{A}, \bar{\Theta}^{\dot{A}})\},
\end{equation}

Calculating the canonical momenta
\begin{equation}\label{momenta}
{\cal P}_{{\cal M}} = { \partial L \over \partial \dot{q}^{{\cal M}} }
= (P_{A\dot{A}}, Z_{AB},
\bar{Z}_{\dot{A}\dot{B}};  P^A, \bar{P}^{\dot{A}};
\pi^{A}, \bar{\pi}^{\dot{A}}) ,
\end{equation}
we obtain the following set of the
primary constraints

\begin{equation}\label{Phi1}
\Phi_{A\dot{B}} \equiv
P_{A\dot{B}} - \lambda_{A} \bar{\lambda}_{\dot{B}} = 0 ,
\end{equation}
\begin{equation}\label{Phi2}
\Phi_{A{B}} \equiv
Z_{A{B}} - \lambda_{A} {\lambda}_{{B}} = 0 ,
\end{equation}
\begin{equation}\label{Phi3}
\Phi_{\dot{A}\dot{B}} \equiv
\bar{Z}_{\dot{A}\dot{B}} - \bar{\lambda}_{\dot{A}} \bar{\lambda}_{\dot{B}} = 0 ,
\end{equation}
\begin{equation}\label{PA=0}
P_{A}=0, \qquad
\bar{P}_{\dot{A}}  = 0 ,
\end{equation}
                  
\begin{equation}\label{DA=0}
D_{A}\equiv - \pi_A + i P_{A\dot{B}} \bar{\Theta}^{\dot{B}} + i
Z_{AB} \Theta^B=0, \qquad
\end{equation}
\begin{equation}\label{bDA=0}
\bar{D}_{\dot{A}}\equiv  \bar{\pi} _{\dot{A}} - i \Theta^B  P_{B\dot{A}}
- i \bar{Z}_{\dot{A}\dot{B}} \bar{\Theta}^{\dot{B}} = 0.
\end{equation}

Because the action \p{action} is invariant under the world line
reparametrization, the canonical Hamiltonian vanishes
 \begin{equation}\label{H=0}
H \equiv \dot{q}^{{\cal M}}{{\cal P}}_{{\cal M}} - L (q^{{\cal M}} ,
\dot{q}^{{\cal M}}) = 0
\end{equation}
It can be deduced that  the set \p{Phi1}-\p{bDA=0} of $14$
bosonic and $4$ fermionic constraints contains
$6$ bosonic and $3$ fermionic first class constraints
\begin{equation}\label{B1}
 B_1 =
 {\l}^{A} {\bar{\l}}^{\dot{B}}
 P_{A\dot{B}}   = 0,
\end{equation}
\begin{equation}\label{B2}
 B_2 =
 {\l}^{A} \hat{\bar{\mu}}^{\dot{B}}
 P_{A\dot{B}}  - \l^A \hat{\mu}^B Z_{AB} = 0,
\end{equation}
\begin{equation}\label{B3}
 B_3 \equiv (B_2)^*=
 \hat{\mu}^{A} {\bar{\l}}^{\dot{B}}
 P_{A\dot{B}}  - \bar{\l}^{\dot{A}} \hat{\bar{\mu}}^{\dot{B}}
 \bar{Z}_{\dot{A}\dot{B}} = 0,
\end{equation}
\begin{equation}\label{B4}
 B_4 =
2 \hat{\mu}^{A} \hat{\bar{\mu}}^{\dot{B}}
 P_{A\dot{B}}  - \hat{\mu}^A \hat{\mu}^B Z_{AB}
 - \hat{\bar{\mu}}^{\dot{A}} \hat{\bar{\mu}}^{\dot{B}} \bar{Z}_{\dot{A}\dot{B}}
  = 0,
\end{equation}
\begin{equation}\label{B5}
 B_5 =
 {\l}^{A} {\bar{\l}}^{{B}}
 Z_{AB} = 0,
\end{equation}
\begin{equation}\label{B6}
 B_6 \equiv (B_5)^*=
 \bar{\l}^{\dot{A}} {\bar{\l}}^{\dot{B}}
 \bar{Z}_{\dot{A}\dot{B}} = 0,
\end{equation}
\begin{equation}\label{F1}
 F_1 =
 {\l}^{A}
 D_{A} = 0,
\end{equation}
\begin{equation}\label{F2}
 F_2 \equiv (F_1)^*=
 \bar{\l}^{\dot{A}} {\bar{D}}_{\dot{A}}= 0,
\end{equation}
\begin{equation}\label{F3}
 F_3 =
 \hat{\mu}^{A}
 D_{A} +
 \hat{\bar{\mu}}^{\dot{A}} {\bar{D}}_{\dot{A}}= 0,
\end{equation}
where we assume that $\l^A \mu_A \not= 0$ and
\begin{equation}\label{hatmu}
 \hat{\mu}^{A}= {\mu^A \over \l^B \mu_B} , \qquad
 \hat{\bar{\mu}}^{\dot{A}} =
{ \bar{\mu}^{\dot{A}} \over \l^{\dot{B}}
{\mu}_{\dot{B}}},
\end{equation}
i.e. $\l^A \hat{\mu}_A = \bar{\l}^{\dot{A}} \hat{\bar{\mu}} =1$.
One can show
\footnote{
We recall  \cite{Dirac} that the first
class constraints are defined as
those whose
Poisson brackets with all constraints weakly vanish.
Then one can show \cite{Dirac}  that the first class constraints  form
the closed
algebra. }
that our
first class constraints \p{B1} - \p{F3}
can be chosen  for any particular form of the
second spinor $\mu^A$ as a function of canonical
variables $(q^{\cal M}, {\cal P}_{{\cal M}})$.
Further we shall propose and motivate the choice
for $\mu^A, ~\bar{\mu}^{\dot{A}}$.

The remaining
$8$ bosonic and $1$ fermionic constraints are the second class ones.
They are
\begin{equation}\label{SB12}
 {\l}^{A} \hat{\bar{\mu}}^{\dot{B}}
 P_{A\dot{B}}  + \l^A \hat{\mu}^B Z_{AB} = 0,  \qquad
 \hat{\mu}^{A} {\bar{\l}}^{\dot{B}}
 P_{A\dot{B}}  + \bar{\l}^{\dot{A}} \hat{\bar{\mu}}^{\dot{B}}
 \bar{Z}_{\dot{A}\dot{B}} = 0,
\end{equation}
\begin{equation}\label{SB34}
 \hat{\mu}^A \hat{\mu}^B Z_{AB} - 1 = 0, \qquad
\hat{\bar{\mu}}^{\dot{A}} \hat{\bar{\mu}}^{\dot{B}}
\bar{Z}_{\dot{A}\dot{B}} -1
  = 0,
\end{equation}
\begin{equation}\label{SB56}
 P_{A} =0, \qquad
 \bar{P}_{\dot{A}} = 0,
\end{equation}

\begin{equation}\label{SF1}
 S_F \equiv \hat{\mu}^{A}
 D_{A} -
 \hat{\bar{\mu}}^{\dot{A}} {\bar{D}}_{\dot{A}}= 0,
\end{equation}
 We see that the number $\#$ of on-shell phase space degrees of freedom
 in our model is
\begin{equation}
 \label{number}
\# = (28_B +8_F) - 2 \times (6_B + 3_F) - (8_B + 1_F) = 8_B + 1_F
\end{equation}
in distinction with the standard massless superparticle model of
Brink--Schwarz
\cite{BS} or Ferber-Shirafuji \cite{F78,S83} containing $6_B + 2_F$
physical degrees of freedom.

In order to explain the difference
in the number of fermionic constraints, let us write down the matrices of
Poisson brackets
for the fermionic constraints \p{DA=0}, \p{bDA=0}.
 In our case it has the form
\begin{equation}
 \label{CDD}
   C_{\a\b} = \left( \matrix{
    \{ D_{A}, D_{B} \}_P  &  \{ D_{A}, \bar{D}_{\dot{B}} \}_P   \cr
     \{ \bar{D}_{\dot{A}}, D_{{B}} \}_P &
 \{ \bar{D}_{\dot{A}}, \bar{D}_{\dot{B}} \}_P \cr }\right)
  =  \left( \matrix{
    \l_{A}\l_{B} &  \l_{A} \bar{\l}_{\dot{B}}   \cr
     \bar{\l}_{\dot{A}} \l_{{B}}  &
 \bar{\l}_{\dot{A}} \bar{\l}_{\dot{B}} \cr }\right)
\end{equation}
while for the standard FS model \cite{F78,S83}
we obtain
\begin{equation}
 \label{FSCDD}
   C^{FS}_{\a\b}
  =  \left( \matrix{
            0 &  \l_{A} \bar{\l}_{\dot{B}}   \cr
     \bar{\l}_{\dot{A}} \l_{{B}}  & 0
 \cr }\right)
\end{equation}
Now it is evident that in our case the rank of the matrix $C$ is one,
while for FS model it is equal to two
$$
rank (C) = 1, \qquad
  rank( C^{FS}) =2.
$$
Consequently,  in our model  there are three fermionic first class constraints
generating three $\kappa$--symmetries \cite{AL}, one more than in the FS model.

\bigskip

In order to clarify the meaning of the superparticle model \p{action}
and present an explicit representation for its physical degrees of freedom,
we shall demonstrate that it admits the supertwistor representation
in terms of independent bosonic spinor $\l^A$, bosonic spinor $\mu^A$
being composed of $\l^A$ and superspace variables
\begin{equation}
 \label{muXZ2}
 \mu^{A} =
 \left(  X^{A\dot{B}}
 + i \Theta^{A} \bar{\Theta}^{\dot{B}}
 \right)
  \bar{\lambda}{}_{{\dot{B}}}
+ 2 z^{AB} \lambda^{B}
+ i \Theta_{A} (\Theta^B
  \lambda_B ), 
\end{equation}
\begin{equation}
 \label{bmuXZ2}
 \bar{\mu}^{\dot{A}} =
 \left(  X^{B\dot{A}}
 - i \Theta^{B} \bar{\Theta}^{\dot{A}}
 \right)
  \lambda_{B}
+ 2\bar{z}^{\dot{A}\dot{B}} \bar{\lambda}_{\dot{B}}
- i \bar{\Theta}^{\dot{A}}
\bar{\Theta}^{\dot{B}}
  \bar{\lambda}_{\dot{B}}
\end{equation}
 and one real  fermionic composite Grassmann variable $\zeta$
\begin{equation}
 \label{zetaTh2}
\zeta = \Theta^A
{\lambda}_{A}
+
\bar{\Theta}^{\dot{A}}
  \bar{\lambda}_{\dot{A}}
\end{equation}
 Eqs. \p{muXZ2} -\p{zetaTh2} describe  $OSp(8|1)$--supersymmetric
 generalization
 of the Penrose correspondence which is alternative
 to the previously known $SU(2,2|1)$ correspondence, firstly proposed by Ferber \cite{F78}.
 Performing integration by parts and neglecting boundary terms we
can express  our action \p{action}  in terms of
$OSp(8|1)$ supertwistor variables as follows:
\begin{equation}\label{actiontw}
 S = - \int  \left(
 {\mu}{}^{A}
d\lambda_{A}
  + \bar{\mu}{}^{\dot{A}} \,
 d{\bar{\lambda}}_{\dot{A}}
  + i d{\zeta} ~{\zeta} \right) \, .
\end{equation}
Eq. \p{actiontw} presents  the free $OSp(8|1)$  supertwistor
action.
 It can be rewritten in the form \p{act0}
with real coordinates
$Y^{A} = ( \mu^\a , \l^\a, \zeta)$ where
real Majorana spinors $\mu^\a, \l^\a$ are obtained from the
Weyl spinors
$(\mu^A, \bar{\mu}^{\dot{A}})$, $(\l^A, \bar{\l}^{\dot{A}})$
by a linear transformation changing
for the $D=4$ Dirac matrices
the complex Weyl
to real Majorana representation.

The action \p{actiontw}
produces only the second class constraints
\begin{equation}
 \label{StwB12}
P^{(\l )}_A - \mu_{A}= 0, \qquad  P^{(\mu )}_A = 0,
\end{equation}
\begin{equation}
 \label{StwB34}
\bar{P}^{(\l )}_{\dot{A}} - \mu_{\dot{A}}= 0, \qquad
\bar{P}^{(\mu )}_{\dot{A}} = 0,
\end{equation}
\begin{equation}
 \label{StwF}
\pi^{(\zeta ) } = i \zeta
\end{equation}
The Dirac brackets for the $OSp(8|1)$ supertwistor coordinates are
\begin{equation}
 \label{PBtwB}
[ \mu_{A},\l^B  ]_D = \d_A^{~B},
\qquad [ \bar{\mu}_{\dot{A}},\bar{\l}^{\dot{B}}  ]_D =
\d_{\dot{A}}^{~\dot{B}},
\end{equation}
\begin{equation}
 \label{PBtwF}
\{ \zeta , \zeta \}_D = - {i \over 2}
\end{equation}
They can be also obtained  after the analysis of
the Hamiltonian system described by the original action \p{action}.
For this result one should firstly perform  gauge fixing for all the gauge symmetries,
 arriving at the dynamical  system which contains only second class 
constraints,
 and then pass to the Dirac brackets
in a proper way (see \cite{SorG} for corresponding analysis
of the BS superparticle model).
This means that the generalization of the Penrose correspondence
\p{muXZ2}, \p{bmuXZ2}, \p{zetaTh2} should be regarded as coming from the
 second class constraints (primary and obtained from the gauge fixing)
 of the original system and, thus,
should be considered as a relations hold in the strong sense
(i.e. as operator identities after quantization) \cite{Dirac}.
Hence, after the quantization performed in the frame of supertwistor approach,
the generalized Penrose relations \p{muXZ2}, \p{bmuXZ2}, \p{zetaTh2}
can be substituted into the wave function
in order to obtain
the $D=4$ superspace description of our quantum system.

\bigskip

We shall discuss now the relation of Eq. \p{muXZ2}, \p{bmuXZ2}, \p{zetaTh2},
\p{actiontw} with the known FS $SU(2,2|1)$ supertwistor description
of the BS superparticle \cite{F78,S83,BBCL,stv,SorG,Town91}.
The standard FS description is given by the action
 \begin{equation}\label{actionst}
 S=  - \int  \left(
 {\mu}{}^{A}
d\lambda_{A}
  + \bar{\mu}{}^{\dot{A}} \,
 d{\bar{\lambda}}_{\dot{A}}
  + i d{\xi} ~\bar{\xi} \right) 
\end{equation}
supplemented by the first class constraint
\begin{equation}\label{U(1)}
 {\mu}{}^{A} \lambda_{A} -
  \bar{\mu}{}^{\dot{A}} \,
 {\bar{\lambda}}_{\dot{A}}
+ 2i \xi
\bar{\xi} = 0
\end{equation}

The $SU(2,2|1)$ supertwistor
$(\l^A, \bar{\mu}_{\dot{A}}, \bar{\xi } )$,
contains complex Grassmann variable $\xi$
and the supersymmetric  Penrose--Ferber correspondence
is given by
 \begin{equation}\label{muX}
 \bar{\mu}^{\dot{A}} =
 \left(  X^{B\dot{A}}
 - i \Theta^{B} \bar{\Theta}^{\dot{A}}
 \right)
  \lambda_{B}
\end{equation}
 \begin{equation}\label{xiTh}
\xi = \Theta^A
{\lambda}_{A}, \qquad \bar{\xi}=
\bar{\Theta}^{\dot{A}}
  \bar{\lambda}_{\dot{A}}.
 \end{equation}
  
 Comparing Eqs. \p{actionst} -- \p{xiTh}  with our $OSp(8|1)$ supertwistor description
 \p{muXZ2} -- \p{actiontw}
 of the superparticle \p{action} with additional central charge coordinates,
 we note that
 \begin{itemize}
 \item
Besides  additional terms proportional to
tensorial central charge coordinates $z^{AB}$, $\bar{z}^{\dot{A}\dot{B}}$,
there is present in \p{bmuXZ2} the second term quadratic in Grassmann
variables.
This second term, however, does not contribute to
the invariant $\mu^A \l_A$.
\item
In our model  we get
\begin{equation}\label{notU(1)}
 {\mu}{}^{A} \lambda_{A} -
  \bar{\mu}{}^{\dot{A}} \,
 {\bar{\lambda}}_{\dot{A}}
 = 2\l_A \l_B z^{AB} - 2 \bar{\l}_{\dot{A}}
\bar{\l}_{\dot{B}} \bar{z}^{\dot{A}\dot{B}}
+ 2i \Theta^A \l_A \bar{\Theta}^{\dot{A}}\bar{\l}_{\dot{A}}
\end{equation}
i.e. we do not have additional first class  constraint generating $U(1)$ symmetry
(compare  to \p{U(1)} of the standard supertwistor formulation). Thus our action
\p{actiontw} is not singular in distinction to \p{actionst}, where the
first class constraint \p{U(1)} should be taken into account, e.g. by
introducing it into the
action with Lagrange multiplier \cite{Town91}.

\item
The complex Grassmann variable $\xi$ \p{xiTh} of FS formalism
is replaced in our case
by the real one $\zeta$ \p{zetaTh2}. This difference implies that in our
supertwistor formalism the limit
$z^{AB}\rightarrow 0, ~~
\bar{z}^{\dot{A}\dot{B}}\rightarrow 0$
does not reproduce the standard $SU(2,2|1)$ supertwistor formalism.
Indeed, this is not surprising if we  take into account that,
from algebraic point of view,
$SU(2,2|1)$ is not a subsupergroup of $OSp(8|1)$.

\end{itemize}

\bigskip

The model (2.1) can be slightly generalized as follows
\begin{equation}\label{acZZ}
S=\int d\tau \left(
\l_A \bar{\l}_{\dot{B}} \Pi_\tau^{A\dot{B}} + 
Z \l_A {\l}_{{B}} \Pi_\tau^{A{B}} +  
\bar{Z} \bar{\l}_{\dot{A}} 
\bar{\l}_{\dot{B}} \bar{\Pi}_\tau^{\dot{B}\dot{B}} \right) \qquad ,
\end{equation}
where $Z$, $\bar{Z}$ are complex numerical constants. 
It appears that for all values of $Z \not= 1$ the model \p{acZZ} will 
have only two $\kappa$-symmetries, and only for particular value $Z=1$ 
we obtain three $\kappa$-symmetries. 
The quantization of the model \p{acZZ} is now under consideration 
\cite{ABLS}.

\bigskip

\section{
$D=10$ and $D=11$ models with one fundamental spinor}
\setcounter{equation}{0}

Recently the most general superparticle model associated with
space--time superalgebra \p{QQZ}
 was proposed  by Rudychev and Sezgin \cite{RS}.
Introducing generalized real superspace
$(X^{\alpha\beta}, \Theta^{\alpha})$
they consider  the following action
\begin{equation}\label{actionRS}
S=\int d\tau L = \int d\tau \left(
P_{\alpha\beta}\,  \Pi_\tau^{\alpha\beta} +
{1\over 2}
e_{\alpha\beta}\,  P^{\alpha\gamma} \, C_{\gamma\delta}
\, P^{\delta\beta}
\right)\, ,
\end{equation}
where  $ \Pi_\tau^{\alpha\beta}= \dot{X}^{\alpha\beta}
-  \dot{\theta}^{(\alpha} \theta^{\beta)}$
\ $(\dot{a}\equiv {d a \over
d\tau})$, $C$ is the charge conjugation matrix and
$e_{\alpha\beta}$ is the set of Lagrange multipliers,
generalizing einbein in the action for standard Brink-Schwarz massless
superparticle \cite{BS}.

Generalized mass shell condition, obtained by varying
$e_{\alpha\beta}$ in \p{actionRS}, takes the form
\begin{equation}\label{BPS}
P^{\alpha\gamma} C_{\gamma\delta} P^{\delta\beta} = 0 \, .
\end{equation}

We shall look for $P^{\a\b}$ expressing it as spinor belinears and 
satisfying the generalized mass shell condition \p{BPS}. 
Particular solution is provided by the following extension of our 
representation \p{CPrep} to any dimension $D>4$ with the use of one real 
$D$-dimensional Majorana spinor $\l_\a$ ($\a=1,..., 2^k$, 
$k=4$ for $D=10$, $k=5$ for $D=11$): 
\begin{equation}\label{CPrepD}
 P_{\a\b} = \l_{\a} \l_{\b}, \qquad (\l_\a)^* = \l_\a, \qquad 
 \end{equation}
 where \p{CPrep} is obtained  if $k=2$.
 The expression \p{CPrepD} solves the BPS condition
 $det P_{\a\b}=0$ as well as  more strong Rudychev-Sezgin
 generalized mass shell constraint \p{BPS} valid in the model \p{actionRS} with antisymmetric charge
conjugation matrix $C~$ ($ C_{\a\b} = - C_{\b\a}$).

Using \p{CPrepD} we get the
multidimensional generalization of our action \p{action}
which reads

\begin{equation}\label{actionD}
S = \int_{{\cal M}^1} \l_{\a} \l_{\b} \Pi^{\a\b}
\end{equation}
$$
 \Pi^{\a\b} = dX^{\a\b} - i d\Theta^{(\a}\Theta^{\b )}, \qquad 
$$
$$
\a =1,...,2^k
$$
and for $k=2$ we get the action \p{action}.

The case
$k=4$ can be treated as describing spinorial
$D=10$ massless superparticle model with $126$ composite tensorial central
charges
$Z_{m_1 \ldots m_5}$ (cf. with \cite{HP,ES}). Indeed, using
the basis of antisymmetric products of $D=10$  sigma matrices
we obtain
\begin{equation}\label{PC10dec}
\l_{\a} \l_{\b} \equiv P_{\a\b} =
P_m \s^m_{\a\b}+
Z_{m_1...m_5}
\s^{m_1...m_5}_{\b\a} ,
  \end{equation}
Contraction of this equation with
$\tilde{\s}^{m\a\b}$ produces the expression for momenta in terms
of bosonic spinors
\begin{equation}\label{Pm10}
P_m = {1 \over 16} \l_{\a} \s_m^{\a\b}\l_{\b}    \qquad
\Rightarrow \qquad P_mP^m = 0.
\end{equation}
The mass shell condition $P_mP^m=0$ appears then as
a result of the $D=10$ identity $(\s_m)_{(\a\b} (\s^m)_{\g ) \d}=0$.

\bigskip 

The action \p{actionD} for $k=8$ can be treated as describing a
$0$--superbrane model
in $D=11$ superspace with
$517$ composite tensorial central charge
described by $32$ components of one real Majorana $D=11$ bosonic
spinor.
In distinction to the above case such model
does not produce  a massless superparticle
\footnote{Note, that the $D=11$ Green--Schwarz superparticle model does exist
and was presented in \cite{BT}}.
Indeed, decomposing \p{CPrepD}
in the basis of
products of $D=11$ gamma matrices, one gets
\begin{equation}\label{PC11D}
\l_{\a} \l_{\b} = P_m \G^m_{\a\b}+
Z_{{m}_1{m}_2} \G^{{m}_1{m}_2}
_{\b\a}+ Z_{m_1...m_5}
\G^{m_1...m_5}_{\b\a} ,
\end{equation}
The $D=11$ energy-momentum vector is then given by
\begin{equation}\label{P11lGl}
P_m = {1 \over 32} \l_{\a} \G^{m\a\b}\l_{\b}
\end{equation}
and the $D=11$ mass-shell condition reads
\begin{equation}\label{0P11P11}
M^2 = P_m P^m = {1 \over 1024} (\l \G^{m}\l ) ~(\l \G^{m}\l )
\end{equation}
Using the $D=11$ Fierz identities
one can prove that the mass shell condition acquires the form
\begin{equation}\label{P11P11}
M^2= P_m P^m = 2 ~Z^{mn} Z_{mn} - 32 ^. 5! ~ Z^{m_1...m_5}
Z_{m_1...m_5}
\end{equation}
with $
 Z_{mn} = - {1 \over 64}\l \G_{mn}\l
$,
$~~
 Z_{m_1...m_5} =  {1 \over 32^.5!}\l \G_{m_1...m_5}\l
$.

If we take into consideration that
the equations of motion for our model \p{actionD} imply that
the bosonic spinor $\l^\a$ is constant  ($d\l^\a=0$), we have to conclude that
\p{actionD} with $k=8$ provides the $D=11$ superparticle model
with mass generated dynamically in a way similar to the
tension generating mechanism, studied in
superstring and higher branes in \cite{generation}.

                   \bigskip

Performing the integration by parts we can rewrite the action \p{actionD}
in the $OSp(1|2^k)$ (i.e. $OSp(1|16)$ for $D=10$ and $OSp(1|32)$
for $D=11$) supertwistor $Y^{{\cal A}}= (\mu^\a, \zeta )$ components:
\begin{equation}\label{actiontwD}
S= - \int (\mu^\a d\l_\a + i d\zeta ~\zeta ), \qquad \a = 1,\ldots , 2^k.
\end{equation}
The generalized Penrose--Ferber  correspondence
between real supertwistors and real generalized superspace looks as follows
\begin{equation}\label{muXZD}
\mu^\a = X^{\a\b} \l_\b - i \Theta^\a (\Theta^\b \l_\b), \qquad
\zeta = \Theta^\a \l_\a .
\end{equation}

\section{A set of $D=11$ massless superparticle models with 
conservation of more then $1/2$ target space supersymmetries}

In order to formulate the model we need to describe 
$SO(1,10)/ (SO(1,1) \otimes SO(9) \semiprod K_9 )$ Lorentz harmonic
formalism.

\subsection{${SO(1,10)\over (SO(1,1) \otimes SO(9) \semiprod K_9 }$
spinor moving frame}

The $SO(1,10)$ valued moving frame matrix $u^{\underline{a}}_{\underline{m}}$
splits into two light--like and 9 space--like vectors \cite{sok}
\begin{equation}\label{11Du}
u^{\underline{a}}_{\underline{m}} =
(u^{++}_{\underline{m}}, u^{--}_{\underline{m}}, u^{I}_{\underline{m}})
\qquad \in \qquad SO(1,10)
\end{equation}
$$
\Leftrightarrow \qquad
u^{\underline{a}\underline{m}} u^{\underline{b}}_{\underline{m}} =
\eta^{\underline{a}\underline{b}} \qquad  \Leftrightarrow \qquad
\cases{
u^{++\underline{m}}u^{++}_{\underline{m}}=0, \cr
u^{--\underline{m}}u^{--}_{\underline{m}}=0, \cr
u^{\pm\pm\underline{m}}u^{I}_{\underline{m}}=0, \cr
u^{I\underline{m}}u^{J}_{\underline{m}}= -\d^{IJ}\cr}
$$
where $I=1,...,9$ is $SO(9)$ vector index.

The $Spin(1,10)$ valued spinor moving frame matrix
$v^{\underline{\a}}_{\underline{\mu}}$ representing the same Lorentz rotation
 \begin{equation}\label{11DGinv}
u^{\underline{a}}_{\underline{m}}
\G^{\underline{m}}_{\underline{\mu}\underline{\nu}} =
 v^{~\underline{\a}}_{\underline{\mu}}
\G^{\underline{a}}_{\underline{\a}\underline{\b}}
v^{~\underline{\b}}_{\underline{\nu}},
\end{equation}
 \begin{equation}\label{11DGinv1}
u^{\underline{a}}_{\underline{m}}
\G_{\underline{a}}^{\underline{\a}\underline{\b}} =
 v^{~\underline{\a}}_{\underline{\mu}}
\G_{\underline{a}}^{\underline{\mu}\underline{\nu}}
v^{~\underline{\b}}_{\underline{\nu}},
\end{equation}
splits into two rectangular blocks
\begin{equation}\label{11Dv}
v^{\underline{\a}}_{\underline{\mu}} =
(v_{\underline{\mu}A}^{~+}, v_{\underline{\mu}A}^{~-})
\qquad \in \qquad Spin(1,10)
\end{equation}
where $A=1,...,16$ is $SO(9)$ spinor index and the sign
superscripts denote the $SO(1,1)$ weight of the vector and spinor harmonics.

As the $Spin(1,10)$ transformations keep invariant not only the gamma
matrices \p{11DGinv}, but the $D=11$ charge conjugation matrix as well
 \begin{equation}\label{11DCinv}
v^{~\underline{\a}}_{\underline{\mu}}
C^{\underline{\mu}\underline{\nu}}
v^{~\underline{\b}}_{\underline{\nu}} =
C^{\underline{\a}\underline{\b}} ,
\end{equation}
the spinor harmonics \p{11Dv} are normalized by
 \begin{equation}\label{11Dvnorm}
v^{+\underline{\mu}}_A v^{~-}_{\underline{\mu}B} = -
v^{-\underline{\mu}}_A v^{~+}_{\underline{\mu}B} = -i \d_{AB} , \qquad
v^{-\underline{\mu}}_A v^{~-}_{\underline{\mu}B} = 0 , \qquad
v^{+\underline{\mu}}_A v^{~+}_{\underline{\mu}B} = 0 . \qquad
\end{equation}
Eqs. \p{11Dvnorm} is equivalent to the following decomposition of
$32 \times 32$ unity matrix
\footnote{The appearance of multiplier $i$ in Eqs. \p{11Dvnorm}, \p{11D1d}
is due to the fact that $D=11$ charge conjugation matrix is imaginary for our
choice of notations and signature $\eta^{\underline{a}\underline{b}} =
diag(+1,-1,\ldots, -1)$}
 \begin{equation}\label{11D1d}
\d^{\underline{\mu}}_{\underline{\nu}} =
i v^{~-}_{\underline{\nu}A}
v^{+\underline{\mu}}_A
- i v^{~-}_{\underline{\nu}A}
v^{+\underline{\mu}}_A
\end{equation}

In a suitable $SO(1,1) \otimes SO(9) \semiprod K_9$ invariant representation
for $D=11$ gamma matrices the Eqs. \p{11DGinv} acquire the form
 \begin{equation}\label{11DGinvd}
u^{++}_{\underline{m}}
\G^{\underline{m}}_{\underline{\mu}\underline{\nu}} =
 2 v^{~+}_{\underline{\mu}A} v^{~+}_{\underline{\nu}A}, \qquad
u^{--}_{\underline{m}}
\G^{\underline{m}}_{\underline{\mu}\underline{\nu}} =
 2 v^{~-}_{\underline{\mu}A} v^{~-}_{\underline{\nu}A}, \qquad
u^{I}_{\underline{m}}
\G^{\underline{m}}_{\underline{\mu}\underline{\nu}} =
 2 v^{~+}_{\{ \underline{\mu}| A} \G^I_{AB} v^{~-}_{\underline{\nu}|B]},
\qquad
\end{equation}
(compare e.g.,  with $D=10$ cases from Refs. \cite{gds,ghs,bzst,bpstv}).
The decomposition of the relations \p{11DGinv1} includes, in particular
 \begin{equation}\label{11DGinv1d}
v^{~-}_{\underline{\mu}A}\G_{\underline{m}}
^{\underline{\mu}\underline{\nu}}
v^{~-}_{\underline{\nu}B} = 2 \d_{AB}u^{--}_{\underline{m}} , \qquad
 \qquad
\end{equation}

\subsection{Action for $D=11$ massless superparticle with
tensorial central charge coordinates}

The twistor-like action for $D=11$
massless superparticle with tensorial central charge coordinates 
has the form
\begin{equation}\label{D11h}
S = \int_{{\cal M}^1} P^{++}_{AB} v^{~-}_{A\mu } v^{~-}_{B\nu} \Pi^{\mu\nu}
\end{equation}
with
$$
 \Pi^{\mu\nu} = dX^{\mu\nu} - i d\Theta^{(\mu}\Theta^{\nu )}
$$
and symmetric $SO(9)$ spin-tensor Lagrange multiplier $P^{++}_{AB}$.

The canonical momenta 
\begin{equation}\label{PX}
P_{\mu\nu} = {\partial L \over \partial \dot{X}^{\mu\nu}} = 
P^{++}_{AB} v^{~-}_{A\mu } v^{~-}_{B\nu} 
\end{equation}
evidently satisfy the BPS condition 
$$
det(P_{\mu\nu}) = 0
$$
as well as the more strong Rudychev-Sezgin generalized mass shell constraint
$$
P_{\mu\rho}C^{\rho\s} P_{\s\nu} = 0.
$$

The rank of the matrix $P_{\mu\nu}$ is less or equal to 
$16$, equal in fact to the rank of the matrix $P^{++}_{AB}$.
As we will demonstrate just this rank defines the number of 
preserved target space supersymmetries. 

The variation of the action \p{D11h} with respect to the coordinate fields  
\begin{equation}\label{vD11h}
\d S = \int_{{\cal M}^1} P^{++}_{AB} v^{~-}_{A\mu } 
v^{~-}_{B\nu} (di_\d \Pi^{\mu\nu} - 2 i d\Theta^{(\mu} \d \Theta^{\nu )} ) 
\end{equation}
$$
i_\d \Pi^{\mu\nu} = \d X^{\mu\nu} - i \d \Theta^{(\mu}\Theta^{\nu )}
$$
includes effectively the $\d\Theta^{\mu}$ variation only in the combination 
$$
d\Theta^{\nu} v^{~-}_{A\nu } P^{++}_{AB}
\d\Theta^{\mu} v^{~-}_{A\mu } 
$$
Thus the half of $\Theta$ variations 
$\d\Theta^{\mu} v^{~+}_{A\mu }$ are not involved in the variation 
of action and, therefore, parameterize the $16$ kappa symmetries. 

When $det (P^{++}_{AB}) \not= 0$, the rest $16$ of the $32$ Grassmann 
variations 
$\d\Theta^{\mu} v^{~-}_{A\mu }$ acts effectively and produce 
nontrivial equations of motion
$$
d\Theta^{\nu} v^{~-}_{A\nu } P^{++}_{AB}= 0 , \qquad 
\Rightarrow \qquad d\Theta^{\nu} v^{~-}_{A\nu }=0.
$$
We see that there are only $16$ kappa symmetries in such dynamical system and 
so it describes the BPS state preserving $1/2$ of the $D=11$ target 
space  supersymmetry. 

\bigskip 

We obtain an important particular case of the model (4.22) with  $det (P^{++}_{AB}) 
\not= 0$ 
 when the Lagrange multiplier $P^{++}_{AB}$ is proportional to the unity 
matrix
$P^{++}_{AB}= P^{++} \d_{AB} $. 
Due to the properties \p{11DGinvd} of the Lorentz harmonic, 
the product of spinor harmonics 
$v^{~-}_{A\mu} v^{~-}_{A\nu}$ is proportional to 
the gamma matrix $\G_{m\mu\nu}$, hence it does not 
contain components proportional to $\G^{mn}_{ \mu\nu}$,   
$\G^{mnklp}_{\mu\nu}$. Thus the central charge coordinates disappear 
from the action which in this case can be equivalently rewritten as  
\begin{equation}\label{acD110}
S = {1 \over 32} 
\int_{{\cal M}^1} P_{--} v^{~-}_{A\mu} v^{~-}_{A\nu} \G_m^{\mu\nu} \Pi^m
\end{equation}
$$
 \Pi^{m} = dX^{m} - i d\Theta^{\mu}\G^m_{\mu\nu} \Theta^{\nu }
$$
The formula \p{acD110} provides the twistor-like formulation of the 
action for the 'standard' $D=11$  massless superparticle
(without tensorial central charge coordinates), whose 'standard'
(Brink--Schwarz type) action was proposed recently in Ref. \cite{bt}.

The generic case of nondegenerate $P^{++}_{AB}$ matrix corresponds the model 
with central charge coordinates and half of $32$ space time supersymmetries 
conserved.

\bigskip

The case with the matrix $P^{++}_{AB}$ having the rank 1 can be described by  
$$P^{++}_{AB}= \l^{+}_A \l^{+}_B$$ with one bosonic $SO(16)$ spinor 
$\l^{+}_A$. The action \p{D11h} in this case reduces to  
\begin{equation}\label{D11h1}
S = \int_{{\cal M}^1} (\l^{+}_A v^{~-}_{A\mu }) 
(\l^{+}_A v^{~-}_{B\nu}) \Pi^{\mu\nu} 
\end{equation}
If one denotes $\l^{+}_A v^{~-}_{A\mu }= \l_\mu$, one arrives to the expression 
$S = \int_{{\cal M}^1} \l_{\mu }\l_{\nu} \Pi^{\mu\nu} $
which formally coincides with the action proposed in 
\cite{BL}. But the composite nature of the bosonic spinor  $\l_\mu$
in the action \p{D11h1} results in the relation 
\begin{equation}\label{32Pll}
32 P_m \equiv \l_{\mu }\G^m_{\mu\nu}\l_{\nu} = 
(\l^{+}_A v^{~-}_{A\mu }) 
(\l^{+}_A v^{~-}_{B\nu}) \G_m^{\mu\nu}= \l^{+}_A \l^{+}_A u^{--}_m , 
\end{equation}
where $u^{--}_m$ is a light-like harmonic vector $u^{--m}u^{--}_m=0$. 
Thus $P_m P^m= 0$ and we conclude that \p{D11h1} describes a {\sl 
massless} 
$D=11$ superparticle with 
central charge coordinate in distinction with 
the $D=11$ model described by (3.4) \cite{BL}, 
where, in general, the particle is massive with  mass generated dynamically 
\cite{generation}. 

Nevertheless both the models (3.4) and \p{D11h1} describe BPS configurations 
with preservation of $31/32$ part of the $D=11$ target space supersymmetries. 

Indeed the variation of the action 
\p{D11h1} includes  effectively  
only one Grassmann variation $\d \Theta^{\mu} \l_\mu$  
(with $\l_\mu$ composed 
from harmonic and $SO(16)$ spinor as in \p{D11h1}),  
which remains the same for the action (3.4), where the $\lambda_\mu$ spinor 
is fundamental (see \cite{BL}). 

\bigskip 

The matrix $P^{++}_{AB}$ of the rank $r$, $ 1< r<8$ can be represented as   
\begin{equation}\label{P++lAlB} 
P^{++}_{AB}= \l^{+s}_A \l^{+s}_B , \qquad s=1,...,r, \qquad 1<r<8  . 
\end{equation}
It is easy to see that such a model describes the BPS states 
preserving ${(32-r) \over 32}$ supersymmetries. 

 \section{Final remarks}

We would like to recall that
in the  'M-theoretic' approach (see e.g. \cite{AT,PKT,SorT})
the tensorial central charges $Z_{m_1\ldots m_p}$ are
considered as carried by p-branes.
 Following such treatment, one should interpret e.g.  in $D=4$ central charges
 $Z_{\mu\nu}$ as an indication of presence of $D=4$ supermembrane ($p=2$).
 The relation of our superparticle model with such $D=4$ membrane states is not clear now and
 can be regarded as an interesting subject for further study.
Here we should only guess that there should be some singular point--like limit
of supermembrane, which should keep the nontrivial topological charge
and increase the number of preserved (realized linearly)
$D=4$ target space supersymmetries.
Similar limiting prescription should be possible
e.g. for 5--branes in $D=10, 11$ leading to the $D=10$ and $D=11$
superparticle actions
\p{actionD} with the relation \p{CPrepD}
describing composite tensor charges.

At the end of the paper we proposed a
generalized FS model for $D>4$. 
The straightforward generalization 
provides us with $D=10$ massless superparticle model 
preserving $15/16$ supersymmetries and 
 $D=11$ superparticle model with arbitrary, in general nonvanishing, mass 
generated dynalically \cite{generation}. 
The latter concerves $31/32$ of the target spase supersymmetries. 
Then we introduce spinor harmonics and  formulate  {\sl massless} $D=11$ 
superparticle model 
preserving $1/2$, $17/32$, $18/32$, $\ldots$, $31/32$ supersymmetries 
dependent on the rank of the Lagrange multiplier matrix 
$P^{++}_{AB}$.  
The case with $1/2$ corresponds to nondegenerate matrix 
$P^{++}_{AB}$: $det(P^{++}_{AB}) \not= 0$. 
For the choice $P^{++}_{AB}= \propto \d_{AB}$, the 
dependence on central charge coordinates disappears and we arrive at 
the twistor-like formulation of the usual massless
$D=11$ superparticle proposed recently by Bergshoeff and Townsend. 

It should be also mensioned that the superparticle model invariant under 
superPoincare symmetries with central charges can be obtained as a 
contraction limit of superparticle model defined on the orthosymplectic 
supergroup manifolds. 
The $D=4$ case ($OSp(4|1)$ model) is now under consideration 
\cite{BLS}.

\bigskip

\centerline{{\bf Acknowledgements}}

\bigskip

The authors would like to thank D. Sorokin for useful discussions. 
The authors are grateful for the hospitality at the
Departamento de Fisica Teorica and financial support of the  
Universidad de Valencia (J.L.) 
and at the Dipartimento di Fisica "Galileo Galilei", Universita 
Degli Study di Padova ed INFN, Sezione di Padova 
(I.B.), which permited to complete the present 
paper. 

The work was supported by the Austrian Science Foundation in the form
of the Lise Meitner Fellowship under the Project
 {\bf M472--TPH}, by the INTAS grants {\bf INTAS-96-308, 
INTAS-93-127-EXT} and {\bf KBN} grant {\bf 2P03B13012}.

\end{document}